Effect of aromatic hydrocarbon addition on *in situ* powder-in-tube processed $MgB_2$ tapes

H. Yamada

General Technology Division, Central Japan Railway Company, 1545-33, Ohyama, Komaki, Aichi 485-0801, Japan

M. Hirakawa

Construction Department, Central Japan Railway Company, 2-1-85, Kounan, Minato-ku, Tokyo 108-8204, Japan

H. Kumakura and H. Kitaguchi

Superconducting Materials Center, National Institute for Materials Science, 1-2-1, Sengen, Tsukuba, Ibaraki 305-0047, Japan

We fabricated *in situ* powder-in-tube processed $MgB_2$/Fe tapes using aromatic hydrocarbon of benzene, naphthalene, and thiophene as additives, and investigated the superconducting properties. We found that these aromatic hydrocarbons were very effective for increasing the Jc values. The Jc values of 20mol% benzene-added tapes reached 130A/mm$^2$ at 4.2K and 10T. This value was almost comparable to that of 10mol% SiC-added tapes and about four times higher than that of tapes with no additions. Microstructure analyses suggest that this Jc enhancement is due to both the substitution of carbon for boron in $MgB_2$ and the smaller $MgB_2$ grain size.

The $MgB_2$ superconductor is expected to be applied to practical superconducting wires because its transition temperature, 39K, is much higher than those of conventional metallic superconductors. The lower cost of the raw materials, Mg and B, than that of Nb, is an additional advantage of $MgB_2$. Recently, small coils using $MgB_2$ wires have been produced. In order to evaluate the potential of $MgB_2$ for power applications, wire processing techniques are now being actively developed throughout the world. The most popular method is the so-called *in situ* powder-in-tube (PIT) method, where constituent powder was packed into a metal tube and cold-worked into wires.[1-3] However, it is necessary to increase the upper critical field and the critical current density to develop useful $MgB_2$ superconducting wire.

To improve these properties, various methods have been reported. They include, for example, the ball-milling method[4,5], the addition of nanometer-size SiC powder[6], the use of $MgH_2$ powder instead of Mg powder[7], and the use of nanometer-size Mg powder[8]. The addition of nanometer-size SiC powder is effective to increase $B_{c2}$ and, hence, Jc in high magnetic fields. This Jc improvement should be attributed to the substitution of carbon from SiC for boron in $MgB_2$.[9]

In this letter, we used aromatic hydrocarbons of benzene ($C_6H_6$), naphthalene ($C_{10}H_8$), and thiophene ($C_4H_4S$) as additives and succeeded in significantly improving the superconducting properties of $MgB_2$ tapes.

We fabricated $C_6H_6$-, $C_{10}H_8$-, and $C_4H_4S$-added $MgB_2$ tapes by the *in situ* PIT method. Commercial $MgH_2$ powder, commercial amorphous B powder, and 10-20mol% $C_6H_6$, 10mol% $C_{10}H_8$, or 10mol% $C_4H_4S$ were mixed. In order to avoid the evaporation or sublimation of the added aromatic hydrocarbons, the powders were mixed for 1 hour using the ball-milling method. The milling was carried out by using a pot and balls made of tungsten-carbide under a high-purity argon gas atmosphere. The mixed powders were put into pure Fe tubes with 6mm in outer diameter and 3.5mm in inner diameter and cold-rolled into tapes with 4mm in width and 0.5mm in thickness using groove-rolling and flat-rolling machines. The mixing of powders and packing of the mixed powder into Fe tubes were carried out under a high-purity argon gas atmosphere by using a glove box in order to avoid the oxidation of the powders. Heat treatment was carried out at 600°C for 1 hour under a flowing argon gas atmosphere. As a reference, 10mol% SiC-added $MgB_2$ tapes with the same size were also fabricated.

The transport critical current, Ic, was measured by a standard four-probe resistive method at 4.2K in magnetic fields. Current and voltage leads were directly soldered to the Fe sheaths of the tapes. Ic measurements were carried out at 4.2K in magnetic fields below 12T. The field was applied parallel to the main plane of the tapes. The criterion of Ic definition was 1μV/cm.

Figure 1 summarizes the field dependence of Jc at 4.2K for the $C_6H_6$-, $C_{10}H_8$-, and $C_4H_4S$-added tapes. As a reference, the data of the pure and 10mol% SiC-added tapes are also shown in the figure. The aromatic hydrocarbon-added tapes show much higher Jc than the pure tape. The field dependence of Jc was also improved by the aromatic hydrogen additions, suggesting that $B_{c2}$ was enhanced by the additions. The 20mol% $C_6H_6$-added tape shows the highest Jc of 130A/mm$^2$ at 4.2K and 10T among the aromatic hydrocarbon-added tapes. This Jc was about four times higher than that of pure tape and almost comparable to that of the 10mol% SiC-added tape. In magnetic fields lower than 8T, $C_6H_6$-added tapes show higher Jc than the 10mol% SiC-added tape. However, the field dependence of Jc for aromatic hydrocarbon-added tapes is slightly larger than that of the SiC-added tape. This can be explained by the difference in the carbon substitution level, as discussed later. The results in Fig. 1 suggest that another aromatic hydrocarbon addition is also effective to enhance the Jc-B properties of MgB$_2$ tapes.

Figure 2 shows the X-ray diffraction (XRD) pattern of the 10mol% $C_6H_6$-added tape. This XRD pattern reveals that MgB$_2$ was obtained as the nearly single phase. We calculated the a-axis and c-axis lengths of the crystalline MgB$_2$ core from the (110) and (002) x-ray diffraction peaks, respectively.

Figures 3(a) and (b) show the a- and c-axis lengths as a function of the additive content. As a reference, the a- and c-axis lengths of the tape prepared with the SiC powder addition are also included in the figure. It was reported that, for the SiC-added tape, the a-axis length shrank, while the c-axis did not change due to the substitution of the carbon from SiC powder for boron.[9] Similar changes of the a- and c-axis lengths were also observed for $C_6H_6$-, $C_{10}H_8$-, and $C_4H_4S$-added tapes, although the decrease of the a-axis length is almost independent of the amount of $C_6H_6$. This result indicates that the substitution of carbon for boron also occurs in the aromatic hydrocarbon-added tapes. Figure 3(a) indicates that the a-axis length of the 10mol% $C_6H_6$-added tape was almost comparable to that of the tape prepared with the 2-3mol% SiC powder addition.

M. Avdeev et al. summarized the a-axis of several MgB$_2$ samples as a function of carbon substitution, x, in Mg(B$_{1-x}$C$_x$)$_2$.[10] Based on these a-axis vs. x curves, the amounts of substituted carbon x in our samples were estimated to be ~0.015 and ~0.03 for the aromatic hydrocarbon-added tapes and 10mol% SiC-added tape, respectively. This smaller amount of carbon substitution for aromatic hydrocarbon additions is the reason that the field dependence of Jc for the aromatic hydrocarbon-added tapes is slightly larger than that of the 10mol% SiC-added tape, as shown in Fig. 1. This amount of substituted carbon for aromatic hydrocarbon-added tapes was much smaller than the 10-20mol% of aromatic hydrocarbon added to the starting powder. Thus, most of the added aromatic hydrocarbons seemed to be vaporized from both ends of the tape during the heat treatment.

Figure 4 shows a comparison of scanning electron micrographs of the fractured surfaces of the $MgB_2$ cores removed from the (a) pure and (b) 10mol% $C_6H_6$-added tapes. Both $MgB_2$ cores show a similar microstructure; however, the $MgB_2$ grain size of the tapes prepared with the $C_6H_6$ addition is smaller than that of tapes with no additions.

In conclusion, the addition of an aromatic hydrocarbon, such as $C_6H_6$, $C_{10}H_8$, and $C_4H_4S$ to the starting powder mixture was effective to increase the Jc values of *in situ*-processed $MgB_2$ tapes. The Jc value of the tapes prepared with the 20mol% $C_6H_6$ addition reached 130A/mm$^2$ at 4.2K and 10T. This Jc value was almost comparable to that of the tapes prepared with the 10mol% SiC powder addition and about four times higher than that of tapes with no additions. We consider that there are two factors for the Jc increase observed in the aromatic hydrocarbon-added tapes.

(1) The carbon atoms of $C_6H_6$, $C_{10}H_8$, and $C_4H_4S$ were substituted for boron, and this substitution enhanced $B_{c2}$ and, hence, increased Jc at high fields. This effect is similar to the effect of the addition of nanometer-size SiC powder.

(2) The $MgB_2$ grain size of the tapes prepared with the $C_6H_6$ addition was smaller than that of tapes with no additions. Such a small $MgB_2$ grain size is effective to enhance flux pinning because the grain boundaries of $MgB_2$ represent effective pinning centers.

**Fig. 1.** Magnetic field dependence of Jc at 4.2K for the aromatic hydrocarbon-added $MgB_2$/Fe tapes heat-treated at $600^oC$ for 1h.

**Fig. 2.** X-ray diffraction pattern of the $MgB_2$/Fe tape prepared with the 10mol% benzene addition.

**Fig. 3.** (a) a-axis and (b) c-axis lengths of the $MgB_2$ cores in the aromatic hydrocarbon- and SiC-added $MgB_2$/Fe tapes, respectively. The horizontal axis shows the amount of additives.

**Fig. 4.** Scanning electron micrographs of the fractured surfaces of the $MgB_2$ core removed from the (a) pure and (b) 10mol% benzene-added tapes.

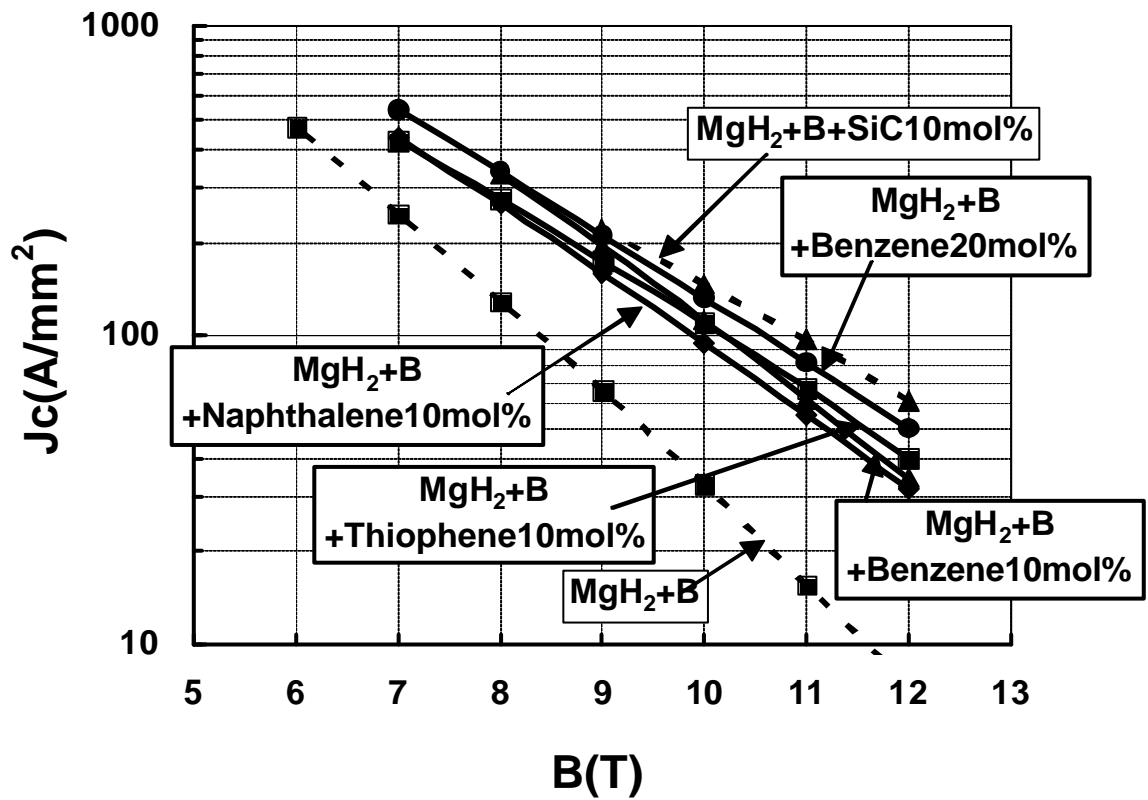

**Fig.1**

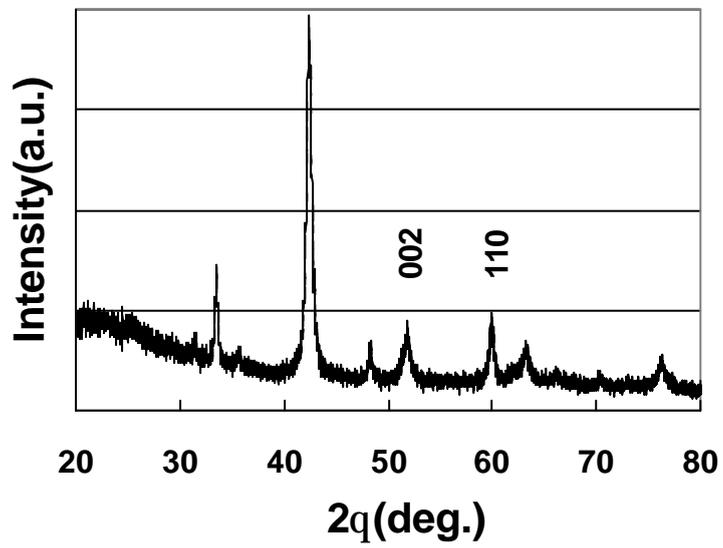

**Fig.2**

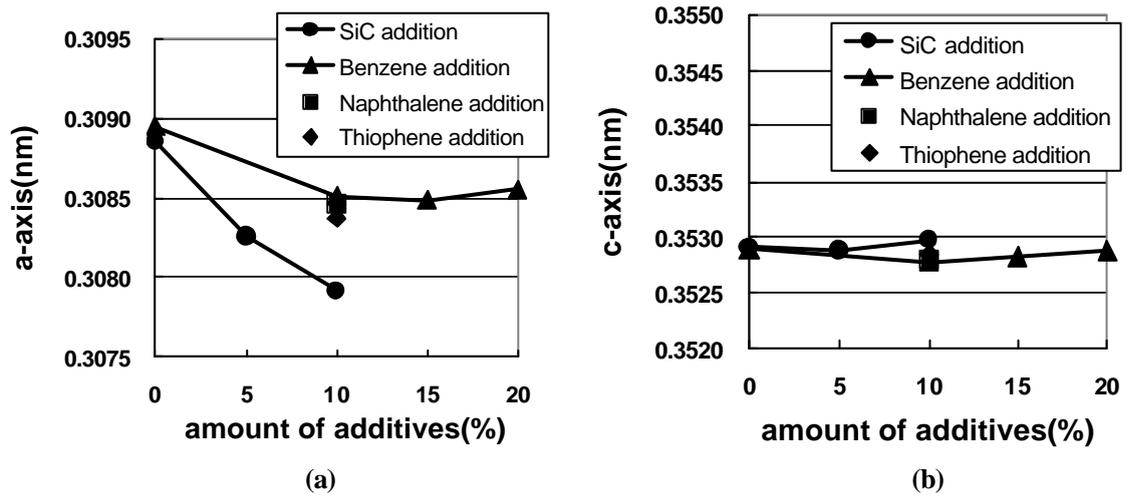

**Fig.3**

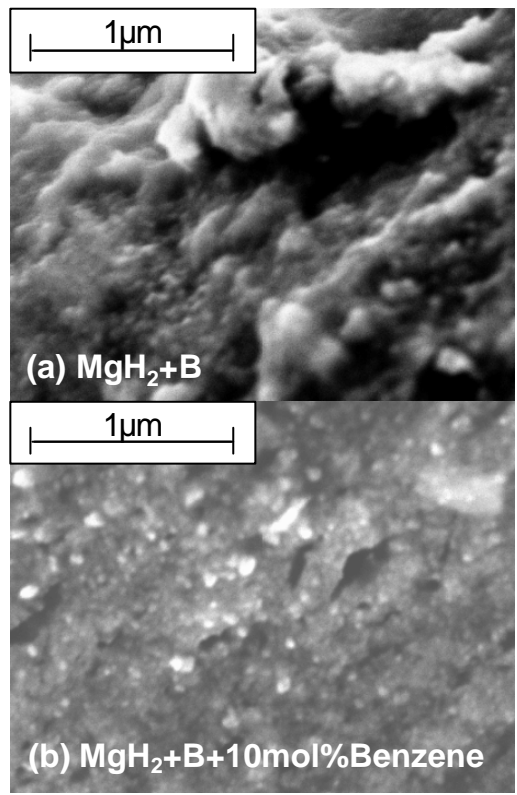

**Fig.4**